\documentclass[aps, prl, superscriptaddress, preprintnumbers, twocolumn, nofootinbib]{revtex4-1}

\usepackage{amsmath}	% AMS Math Package
\usepackage{amsthm}	% Theorem Formatting
\usepackage{amssymb}	% Math symbols such as \mathbb
\usepackage{datetime}
\usepackage{graphicx}
\usepackage{color}
\usepackage{verbatim}

\usepackage[colorlinks=true, citecolor=midblue, linkcolor=midblue, urlcolor=midblue]{hyperref}

\usepackage{amssymb, amsmath} % also includes the command \mathpzc{} for calligraphic letters (uppercase and lowercase) in math mode

\usepackage{bbm} % font for double-lined letters _and_ numbers, e.g. matrix one; usage: \mathbbm{...}
\usepackage{bm} % bold math fonts; usage: \bm{} 

\usepackage[english]{babel}
\usepackage{slashed}
\usepackage{wasysym}
\usepackage{pifont}

\usepackage{charter}

\usepackage{graphicx}
\usepackage{color}

\usepackage{picture,xcolor}

\date{}

\definecolor{grey}{rgb}{0.4,0.4,0.4}
\definecolor{dullmagenta}{rgb}{0.4,0,0.4}
\definecolor{darkblue}{rgb}{0,0,0.4}
\definecolor{midblue}{rgb}{0,0,0.5}
\definecolor{midred}{rgb}{0.5,0,0}
\definecolor{orange}{rgb}{1,0.5,0}
\definecolor{lightbrown}{rgb}{0.75,0.5,0.25}
\definecolor{tan}{cmyk}{0.14,0.42,0.56,0}
\definecolor{djunglegreen}{cmyk}{0.99,0,0.52,0}
\definecolor{lightgreen}{rgb}{0,1,0}
\definecolor{olivegreen}{cmyk}{0.64,0,0.95,0.40}
\definecolor{midgreen}{rgb}{0.0,0.675,0.0}
\definecolor{darkgreen}{rgb}{0,0.5,0}

\newcommand{\be}{\begin{equation}}
\newcommand{\ee}{\end{equation}}

\newcommand{\q}{\quad}

\newcommand{\vs}{\vspace}

\renewcommand{\.}{\hspace{0.5mm}}

\newcommand{\srm}{\ensuremath{\mathrm{s}}}

\newcommand{\Ocal}{\ensuremath{\mathcal{O}}}

\newcommand{\Rcal}{\ensuremath{\mathcal{R}}}

\renewcommand{\d}{\ensuremath{\mathrm{d}}}
\newcommand{\eg}{e.g.}

\let\baraccent=\= % rename builtin command \= to \baraccent
\renewcommand{\=}[1]{\stackrel{#1}{=}} % for putting numbers above =

\theoremstyle{definition}

\theoremstyle{remark}

\settimeformat{ampmtime}

\usepackage{float}

\begin{document}

\title{Waves from the Centre: Probing PBH and other Macroscopic Dark Matter with LISA}

\author{Florian K{\"u}hnel}
\email{florian.kuehnel@physik.uni-muenchen.de}
\affiliation{The Oskar Klein Centre for Cosmoparticle Physics,
	Department of Physics,
	Stockholm University,
	AlbaNova University Centre,
	Roslagstullsbacken 21,
	SE--106\.91 Stockholm,
	Sweden}

\affiliation{Department of Physics,
	School of Engineering Sciences,
	KTH Royal Institute of Technology,
	AlbaNova University Centre,
	Roslagstullsbacken 21,
	SE--106\.91 Stockholm,
	Sweden}

\author{Andrew Matas}
\email{andrew.matas@aei.mpg.de}
\affiliation{Albert-Einstein-Institut,
	Max-Planck-Institut f{\"u}r Gravitationsphysik,
	D-14776 Potsdam-Golm,
	Germany}

\author{Glenn D.~Starkman}
\email{glenn.starkman@case.edu}
\affiliation{CERCA/ISO,
	Department of Physics,
	Case Western Reserve University,
	10900 Euclid Avenue, 
	Cleveland, 
	OH 44106, 
	USA}

\author{Katherine Freese}
\email{ktfreese@umich.edu}
\affiliation{Department of Physics,
	University of Michigan,
	Ann Arbor,
	MI 48109,
	USA}

\affiliation{The Oskar Klein Centre for Cosmoparticle Physics,
	Department of Physics,
	Stockholm University,
	AlbaNova University Centre,
	Roslagstullsbacken 21,
	SE--106\.91 Stockholm,
	Sweden}
	
\affiliation{Theory Group, 
	Department of Physics, 
	University of Texas at Austin, 
	Austin, TX 78712,
	USA}

\date{\formatdate{\day}{\month}{\year}, \currenttime}

\begin{abstract}
A significant fraction of cosmological dark matter can be formed by very dense macroscopic objects, for example primordial black holes. Gravitational waves offer a promising way to probe these kinds of dark-matter candidates, in a parameter space region that is relatively untested by electromagnetic observations. In this work we consider an ensemble of macroscopic dark matter with masses in the range $10^{-13}$ -- $1\ M_{\odot}$ orbiting a super-massive black hole. While the strain produced by an individual dark-matter particle will be very small, gravitational waves emitted by a large number of such objects will add incoherently and produce a stochastic gravitational-wave background. We show that LISA can be a formidable machine for detecting the stochastic background of such objects orbiting the black hole in the centre of the Milky Way, Sgr\.${\rm A}^{\!*}$, if a dark-matter spike of the type originally predicted by Gondolo and Silk forms near the central black hole.\\
\end{abstract}

\maketitle

%%%%%%%%%%%%%%%%%%%%%%%%%%%%%%%%%%%%%%%%%%%%%%%%%%%%%%%%%%%%

%%%%%%%%%%%%%%
% INTRODUCTION
%%%%%%%%%%%%%%

% DM, WIMPS
According to the current standard cosmological model, approximately $25$\% of the energy density of the Universe is in the form of so-called cold dark matter{\,---\,}non-relativistic objects which collectively act as a perfect fluid of negligible pressure. The leading candidates have long been axions \cite{Peccei:1977hh, Weinberg:1977ma, Wilczek:1977pj} as well as weakly-interactive massive particles (WIMPs) heavy ($\gtrsim 1\.$GeV) particles outside the Standard Model of particle physics, possessing very small scattering cross-sections on each other and on Standard-Model particles. For a review of particle dark matter, please see \cite{Jungman:1995df, Bertone:2004pz}.

% Macros
With the continued non-detection of WIMP dark matter, and the failure of long-predicted Beyond the Standard-Model physics to materialise at the Large Hadron Collider, the case for alternative, and especially Standard Model, candidates has grown stronger, and attracted increasing attention. It has also long-been recognised that there are viable dark-matter candidates of much greater mass, notably primordial black holes (PBH) \cite{1967SvA....10..602Z, Carr:1974nx} (see also Refs.~\cite{Dolgov:1992pu, Jedamzik:1996mr, Niemeyer:1997mt, Jedamzik:2000ap, Musco:2004ak, Musco:2008hv, Capela:2012jz, Griest:2013aaa, Belotsky:2014kca, Young:2015kda, Frampton:2015xza, Bird:2016dcv, Kawasaki:2016pql, Carr:2016drx, Kashlinsky:2016sdv, Clesse:2016vqa, Green:2016xgy, Kuhnel:2017pwq, Akrami:2016vrq, Garcia-Bellido:2017fdg, Garcia-Bellido:2017qal, Carr:2017jsz}, and in particular Ref.~\cite{Carr:2019kxo} in which it is shown that the thermal history of the Universe naturally explains the entirety of the dark matter as well as several other cosmic conundra such as the origin of the seeds of supermassive black holes) and objects of nuclear density (\eg~Ref.~\cite{Witten:1984rs, Lynn:1989xb, Kusenko:1997si, Zhitnitsky:2002nr, Lynn:2010uh}), either of which could potentially be the result of Standard-Model physics in the early Universe. For the purposes of this paper, we will refer to all such macroscopic dark-matter candidates, including PBHs, generically as \emph{macros}.

% EM limits on macros
There are well-known limits on macros from microlensing of Milky Way and Magellanic Cloud stars \cite{Allsman:2000kg, Tisserand:2006zx, Carr:2009jm, PhysRevLett.111.181302} limiting the abundance of macros above approximately $4 \times 10^{24}\.{\rm g}$. However, between about $2 \times 10^{17}\.{\rm g}$ and $10^{22}\.{\rm g}$ there is an unconstrained window for anything of approximately ordinary matter density or greater \cite{Jacobs:2014yca}.\footnote{Note that recently several constraints have been revised, such as those from femtolensing \cite{Katz:2018zrn} and well as from microlensing of stars in M31 \cite{Niikura:2017zjd}, leading to significant relaxation of those constraints.} Candidates of approximately nuclear or greater density ($\gtrsim 2.3 \times 10^{17}\.{\rm kg\,m}^{-3}$) are also unconstrained from $55\.{\rm g}$ to $2 \times 10^{17}\.{\rm g}$ \cite{Jacobs:2014yca}, although if, as expected, primordial black holes emit Hawking radiation, then they would have evaporated before now if their masses were below approximately $10^{15}\.{\rm g}$. 

% GWs, LIGO, LISA
With the dawn of gravitational-wave astronomy, it is interesting to use gravitational-wave observations to probe the nature of dark matter. The Advanced Laser Interferometer Gravitational-wave Observatory (LIGO) \cite{aLIGO_2015} and Advanced Virgo \cite{aVirgo_2015} detectors, operating in approximately the $10$ -- $10^{3}\.$Hz band, are now regularly detecting the merger of compact binaries whose total masses are of order a few to tens of solar masses \cite{gw150914, gw151226, o1bbh, gw170608, gw170104, gw170814, gw170817, LIGOScientific:2018mvr}. There has been significant attention to the possibility that multi-solar-mass black holes, such as those detected by Advanced LIGO, could be the dark matter \cite{Bird:2016dcv, Mandic:2016lcn, Kovetz:2016kpi, Sasaki:2016jop, Sasaki:2018dmp}. 
The Laser Interferometer Space Antenna (LISA) \cite{lisa-proposal}, is a space-based gravitational-wave detector which will operate in the $10^{-4}$ -- $10^{-1}$\.Hz band. The detectability of individual macros/objects orbiting Sgr\.${\rm A}^{\!*}$ with LISA has been discussed in Ref.~\cite{Freitag:2002nm}. Sesana \cite{PhysRevLett.116.231102} and Clesse {\it et al.} \cite{Clesse:2016ajp} have suggested that if the dark matter is composed of multi-solar-mass black holes that LISA could detect their merger. Recently, Bartolo {\it et al.} \cite{Bartolo:2018evs, Bartolo:2018rku} have argued that LISA will be able to detect the two- and three-point correlator of gravitational waves originating from the initial gravitational collapse to PBHs of mass around $10^{-22}\.$g.

% Our plan
In this work, we explore whether LISA can explore macros in the whole mass range $10^{-13}$ -- $1\ M_{\odot}$ $( 2 \times 10^{20}\.{\rm g}$ -- $2 \times 10^{33}\.{\rm g})$. Because of its relative proximity, a promising target is Sagittarius ${\rm A}^{\!*}$ (Sgr\.${\rm A}^{\!*}$), a super-massive black hole (SMBH) with a mass of $4 \times 10^{6}\.M_{\odot}$ at the centre of our Galaxy \cite{SagA_params}. Even though the strain associated with the gravitational waves from an individual macro orbiting Sgr\.${\rm A}^{\!*}$ is below the sensitivity threshold of any near-term gravitational-wave detectors such as LISA, the {\it collective} signal from a large number of such macros might be detectable as a stochastic gravitational-wave background.

%%%%%%%%%%%%%%
% Circular orbits
%%%%%%%%%%%%%%
In this work, we aim for a conservative estimate on the gravitational-wave detection prospects of a possible macro dark-matter distribution in our Galaxy. We assume that we can treat the objects non-relativistically, so we use the Peters-Mathews formula \cite{Peters:1963ux} for the time-averaged power emitted by one macro of mass $\mu$ orbiting ${\rm Sgr\.A}^{\!*}$ (with mass $M$) at a distance $a$
\begin{align}
	P
		&=
					\frac{ 32 }{ 5 }\.
					\frac{G^{4}}{c^{5}}\.
					\frac{M^{2} \mu^{2}\.( M + \mu )}
					{a^{5}}
		\approx
					\frac{ 32 }{ 5 }\.
					\frac{q^{2}\.G^{4} M^{5}}{a^{5} c^{5}},
					\label{eq:peters-mathews}
\end{align}
where $q \equiv \mu / M$ is the mass ratio. For simplicity, we assume that all macros have the same eccentricity; specifically, we assume circular orbits.

At sufficiently large macro mass, macros near the black hole will lose energy due to gravitational wave emission and plunge into the central black hole. To account for this, we only include orbits where the timescale for the orbital frequency to change, $\tau = \nu / \dot{ \nu }$, is larger than a cutoff $\tau_{\rm min}$ (note that to within a factor of $\Ocal( 1 )$, the time change of orbital frequency is equal to the time to coalescence.). Here and throughout the text, $\nu$ refers to the emitted gravitational-wave frequency. We estimate the inspiral time using \cite{Maggiore:1900zz}\begin{align}
	\tau
		&\equiv
					\frac{ \nu }{ \dot{ \nu } }
		=
					\frac{ M }{ \mu }\.\frac{ 5 }{ 96\.\pi^{8/3} }\.
					\frac{ c^{5} }{ (G M)^{5 / 3} }\.\nu^{-8/3}
					\;.
					\label{eq:tau-1}
\end{align}
Imposing the condition $\tau>\tau_{\rm min}$ leads to a $\mu$-dependent maximum frequency at which gravitational-waves are emitted. We consider two cases for $\tau_{\rm min}$. First, we consider a highly conservative option $\tau_{\rm min} = H_{0}^{-1}$, which guarantees that the dark-matter profile is stable and the macros do not fall into the black hole due to gravitational wave emission within the age of the universe. Second, we optimistically assume that the dark matter in orbits close to the black hole is refilled efficiently, and take $\tau_{\rm min}$ equal to the observational time of the LISA mission, which we take to be 5 years. In either case, the macros we consider follow very stable orbits and each macro emits gravitational waves at approximately a single frequency.

\begin{figure}
	\vs{-3mm}
	\includegraphics[width=0.49\textwidth]{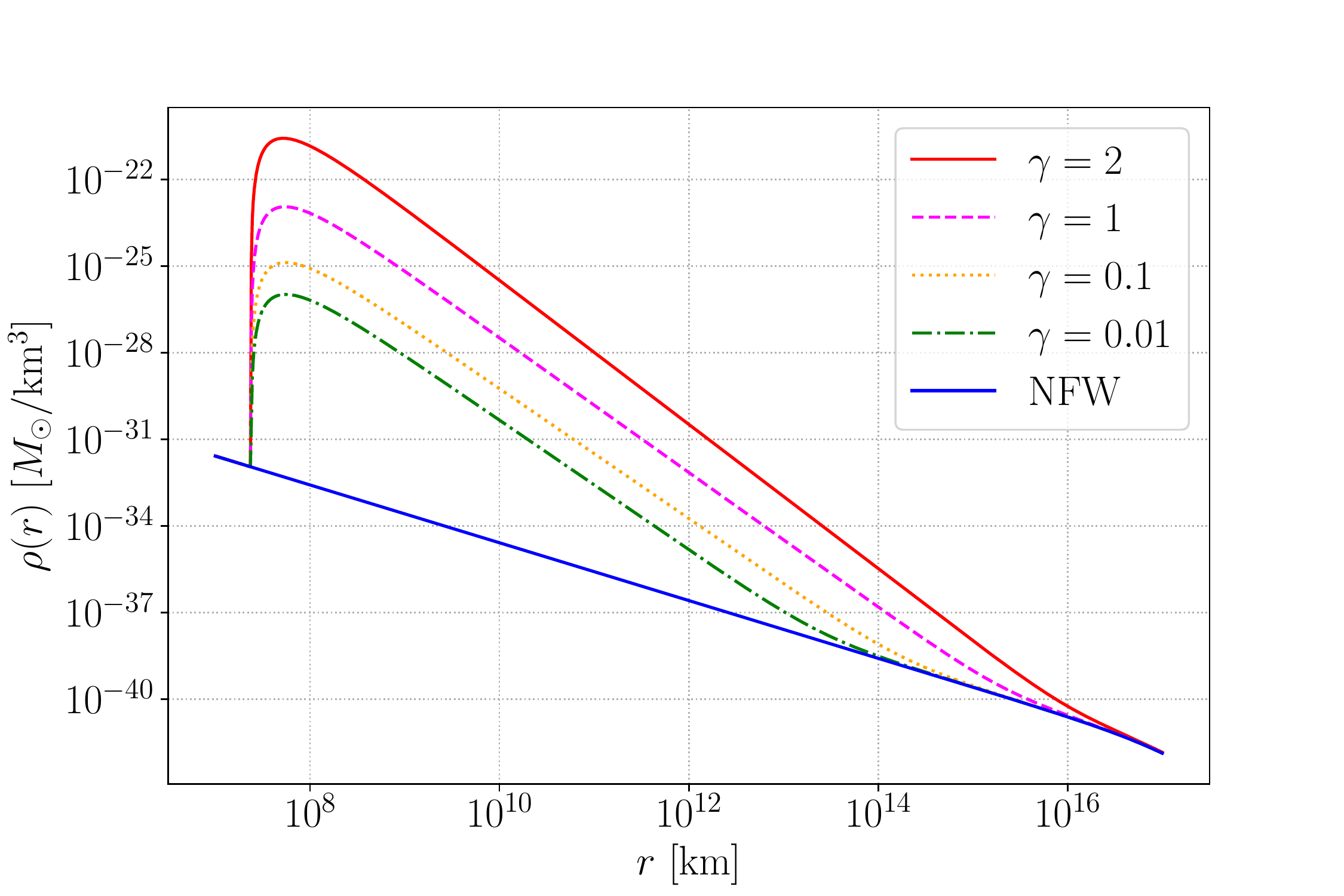}
	\caption{
		Dark-matter density profile as a function of radius. 
		Depicted are a pure NFW profile 
		as well as the onset of the spike density profiles 
		Eq.~\eqref{eq:rho} with $\gamma = 0.1,\.1,\.2$ 
		(bottom to top)
		for $M = M_{{\rm Sgr\mspace{1mu}A}^{\!*}}$.
		}
	\label{fig:rho}
\end{figure}

%%%%%%%%%%%%%%
% Dark-matter spike
%%%%%%%%%%%%%%
The stochastic gravitational-wave signal seen by LISA is dominated by orbits which pass very close to the central black hole. We consider several scenarios for the dark-matter distribution. First, we consider a Navarro-Frenk-White (NFW) profile \cite{Navarro:1995iw}, which is given by
\begin{align}
	\rho_{\rm NFW}( r )
		&=
					\rho_{\srm}\.
					\frac{ r_{\srm} }
					{r
					\left(
						1
						+
						\frac{r}{r_{\srm}}
					\right)^{\!2} }
					\; ,
					\label{eq:NFW}
\end{align}
with the scale radius $r_{\srm} = 24.42\ {\rm kpc}$ and density $\rho_{\srm} = 0.184\.{\rm GeV}\.{\rm cm}^{-3}$ (see Ref.~\cite{Cirelli:2010xx}). Second, we consider that a dark-matter spike can form by adiabatic collapse, following the model of Gondolo and Silk \cite{Gondolo:1999ef}, including relativistic corrections by Sadeghian \emph{et.~al.}~\cite{Sadeghian:2013laa} (see also \cite{1980ApJ...242.1232Y,1986ApJ...301...27B} for the original descriptions of adiabatic collapse). In this scenario, the dark-matter density near the Galactic Centre is enhanced relative to an NFW profile due to adiabatic accretion of dark matter by Sgr\.${\rm A}^{\!*}$. Sadeghian \emph{et.~al.} suggest the following simple analytic approximation for the dark-matter spike density
\begin{align}
	\rho_{\rm sp}( r )
		&\approx
					\left(
						1
						-
						\epsilon
					\right)
					\rho_{R}
					\left(
					 	1
						-
						\frac{2 R_{\rm S}}{ r }
					\right)^{\!3}
					\left(
						\frac{ R_{\rm sp} }{ r }
					\right)^{\!\gamma_{\rm sp}}
					\; ,
					\label{eq:rho}
\end{align}
where $\epsilon = 0.15$, and where $2\.R_{\rm S} < r < R_{\rm sp}$. Here, $R_{\rm S} = 2\.G\.M_{{\rm Sgr\.A}^{\!*}} / c^{2} \simeq 3 \ ( M_{{\rm Sgr\.A}^{\!*}} / M_{\odot} )\.{\rm km}$ is the Schwarzschild radius of Sgr\.${\rm A}^{\!*}$, and $R_{\rm sp} \equiv \alpha_{\gamma}\.r_{0}\.[ M_{{\rm Sgr\.A}^{\!*}} / ( \rho_{0}\.r_{0}^{3} ) ]^{1 / ( 3 - \gamma )}$, where the normalization $\alpha_{\gamma}$ is numerically derived for each power-law index $\gamma$. We define $\rho_{R} \equiv \rho_{0}\.( R_{\rm{sp}} / r_{0} )^{- \gamma}$, where $\gamma_{\rm sp} \equiv ( 9 - 2\.\gamma ) / (4 - \gamma )$ (see Refs.~\cite{Gondolo:1999ef, Nishikawa:2017chy} for details). Fig.~\ref{fig:rho} depicts the radial behaviour of the halo profile \eqref{eq:rho}. We note that this formula assumes that Sgr\.${\rm A}^{\!*}$ has negligible spin. We expect this to be a conservative assumption, as the signal-to-noise ratio is dominated by the smallest orbits, and the innermost stable circular orbit for prograde orbits around a spinning black hole is smaller than for a non-spinning black hole. However, we leave a detailed consideration of the effect of black hole spin for future work. We note that two-body relaxation might diminish the spike. An estimate based on not-yet-published work \cite{Kamionkowski_private} suggests that in the upper portion of the dark-matter-mass range considered here, this may reduce concentration of the central spike. Initial estimates of the effect do not change the primary conclusions of the paper regarding detectability of a signal.

%%%%%%%%%%%%%%
% Omega_GW
%%%%%%%%%%%%%%
The stochastic background is characterised in terms of the energy density per unit logarithmic frequency interval,
\begin{equation}
	\Omega_{\rm GW}(\nu)
		=
				\frac{ \nu }{ \rho_{c} }\.
				\frac{ {\rm d}\rho_{\rm GW}}{{\rm d} \nu}
				\; ,
\end{equation}
where $\rho_{\rm GW}$ is the energy density in gravitational waves, and $\rho_{c} = 3\.H_{0}^{2} / ( 8 \pi G )$ is the critical energy density needed to have a spatially flat universe. We take the value $H_{0}= 67.4\,{\rm km}\.{\rm s}^{-1}\.{\rm Mpc}^{-1}$ for the present-day value of the Hubble parameter as measured by Planck \cite{Aghanim:2018eyx}. We will in turn focus on continuous emission of a number of $N$ macros, all with the same mass $\mu$. The generalisation to an extended mass distribution is straight forward and shall not be given here. For simplicity, since we are only interested in an order of magnitude estimate for the strain power needed for detection, we assume that the macros are isotropically distributed on the sky at the distance of Sgr\.${\rm A}^{\!*}$, and an isotropic distribution of orbital planes. For circular orbits, which emit gravitational waves at a single frequency $\nu$, $\Omega_{\rm GW}$ is given by incoherently summing the power emitted by each macro
\begin{align}
	&\Omega_{\rm GW}( \nu )
		=
					\sum_{k} \frac{\nu P_{k}}{4\pi\.c\.\rho_{c}\.r^{2}}\,
					\delta(
						\nu
						-
						\nu_{k}
					)\.
					\theta(\nu_{\rm max}-\nu)
					\notag\\[1mm]
		&\q\approx
					\frac{ 512\.\pi }{ 45\.H_{0}^{2}\.c^{8} }\.\frac{ \mu^{2} }{ M^{2} }\.
					\frac{ ( G M )^{9/2} }{r^{2}\.a^{1 / 2}\.\mu}\,
					\nu\,\rho_{\rm macro}( a )\.
					\theta(\nu_{\rm max}-\nu)
					\; ,
					\label{eq:}
\end{align}
where in the first line the index $k$ labels the macros, the power $P_{k}$ is given by Eq.~\eqref{eq:peters-mathews}, and to obtain the second line we have replaced a sum over individual macros with an integral over the macro density and have done the integral. The frequency cutoff $\nu_{\rm max}$ is defined as $\nu_{\rm max} = {\rm min}(\nu_{\rm plunge},\nu_{\rho})$, where $\nu_{\rm plunge}$ is found by inverting Equation~\ref{eq:tau-1} with $\tau = \tau_{\rm min}$, and $\nu_{\rho}$ is the gravitational wave frequency emitted by an orbit with radius $2\.R_{{\rm S}}$, the minimum radius of the dark-matter spike in Eq~\ref{eq:rho}. The theta function $\theta(\nu_{\rm max}-\nu)$ thus imposes the constraints that the orbits contributing to $\Omega_{\rm GW}$ should be included in the density distribution, Eq.~\eqref{eq:rho}, and also should not plunge into the black hole due to GW emission on a timescale shorter than $\tau_{\rm min}$. The orbital radius $a$ is related to the emitted gravitational wave frequency $\nu$ by $a \equiv ( G M )^{1 / 3} / ( \pi \nu )^{2 / 3}$, and $r$ is the distance from Earth to the galactic centre. The quantity $\rho_{\rm macro}( a )$ is the macro mass density, which will be assumed to follow the dark-matter distribution, $\rho_{\rm macro} = f_{\rm macro}\.\rho_{\rm sp}$, where $f_{\rm macro}$ is the fraction of dark matter residing in macros. 

\begin{figure}[t]
	\includegraphics[width=0.49\textwidth]{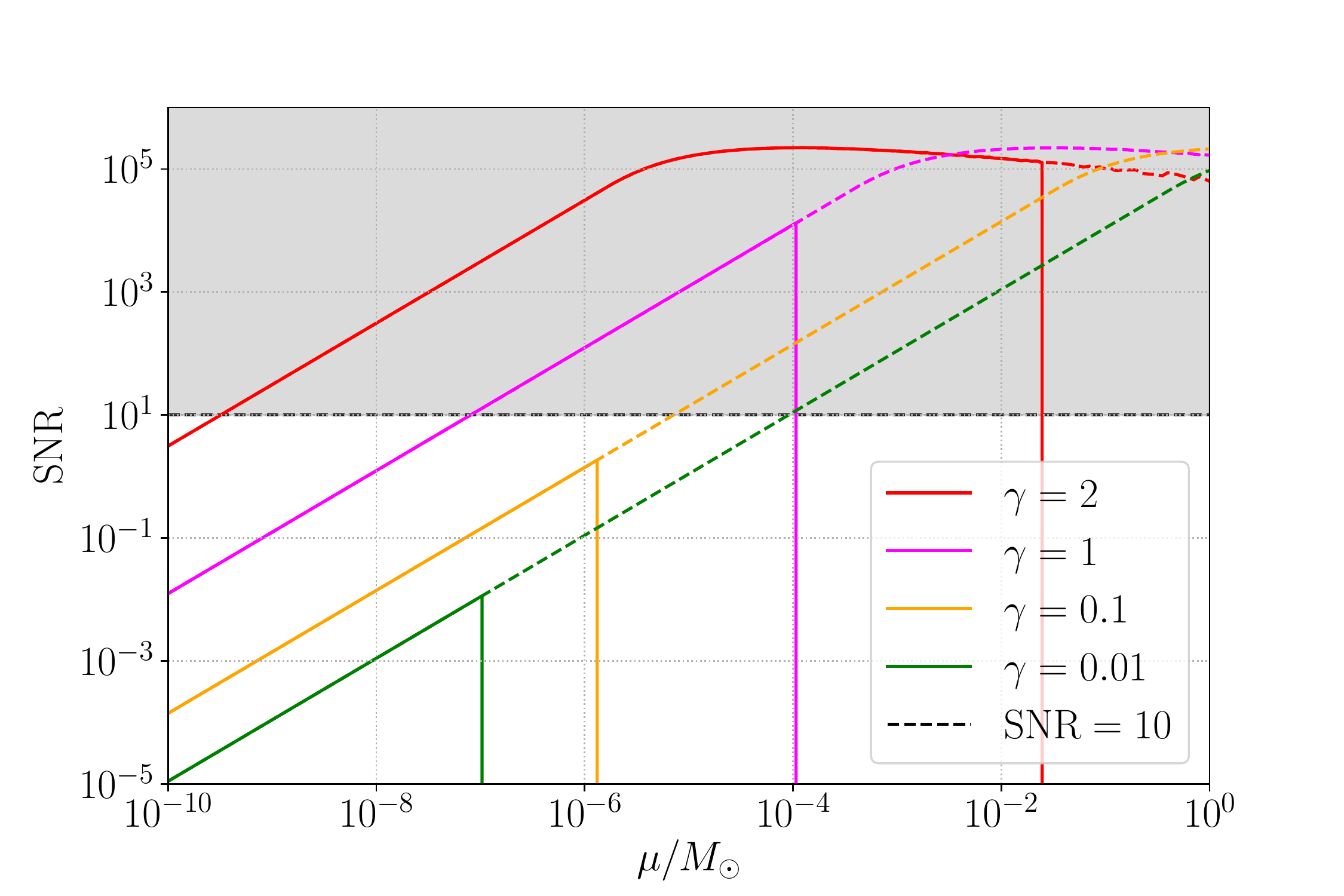}
	\vs{7mm}
	\includegraphics[width=0.49\textwidth]{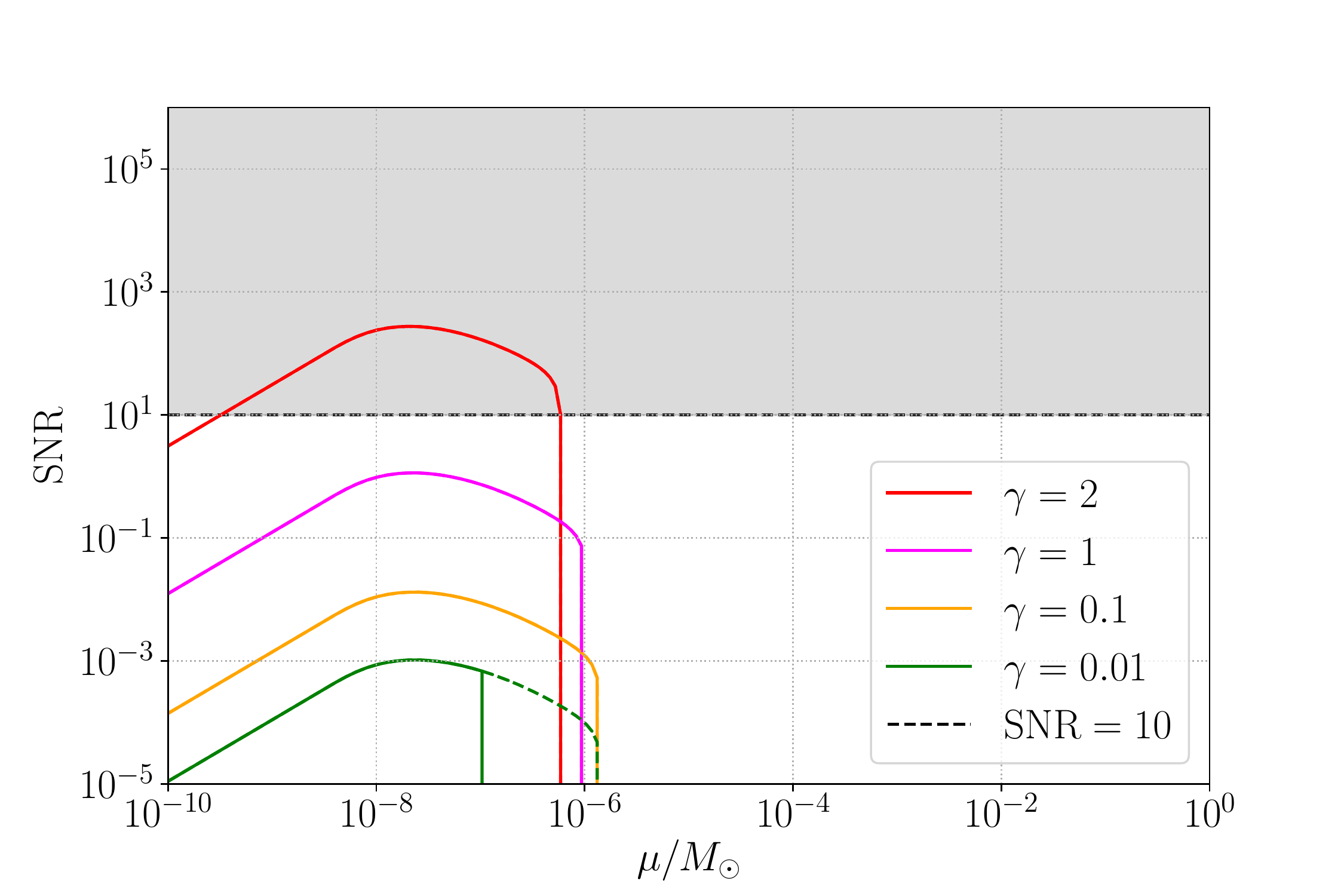}
	\caption{
		Signal-to-noise ratio (SNR) for circular orbits macros of mass $\mu$
		around Sgr\.${\rm A}^{\!*}$. 
		The underlying dark-matter distribution
		is the same as in Fig.~\ref{fig:rho}.
		The upper panel shows results for $\tau_{\rm min} = 5\.{\rm years}$, 
		while the lower panel utilises $\tau_{\rm min} = H_{0}^{-1}$.
		The gray band shows the region that can be detected with ${\rm SNR}=10.$
		In this figure, we assume the dark-matter fraction is 
		$f_{\rm macro} \equiv \Omega_{\rm macro} / \Omega_{\rm DM} = 1.$
		The solid lines represent regions where the signal is expected to be stochastic, 
		while dashed lines represent an extrapolation of the stochastic sensitivity into a parameter
		space where individual sources may be resolvable.
		\vs{2mm}
		}
	\label{fig:SNR}
\end{figure}

\begin{figure}[t]
	\includegraphics[width=0.49\textwidth]{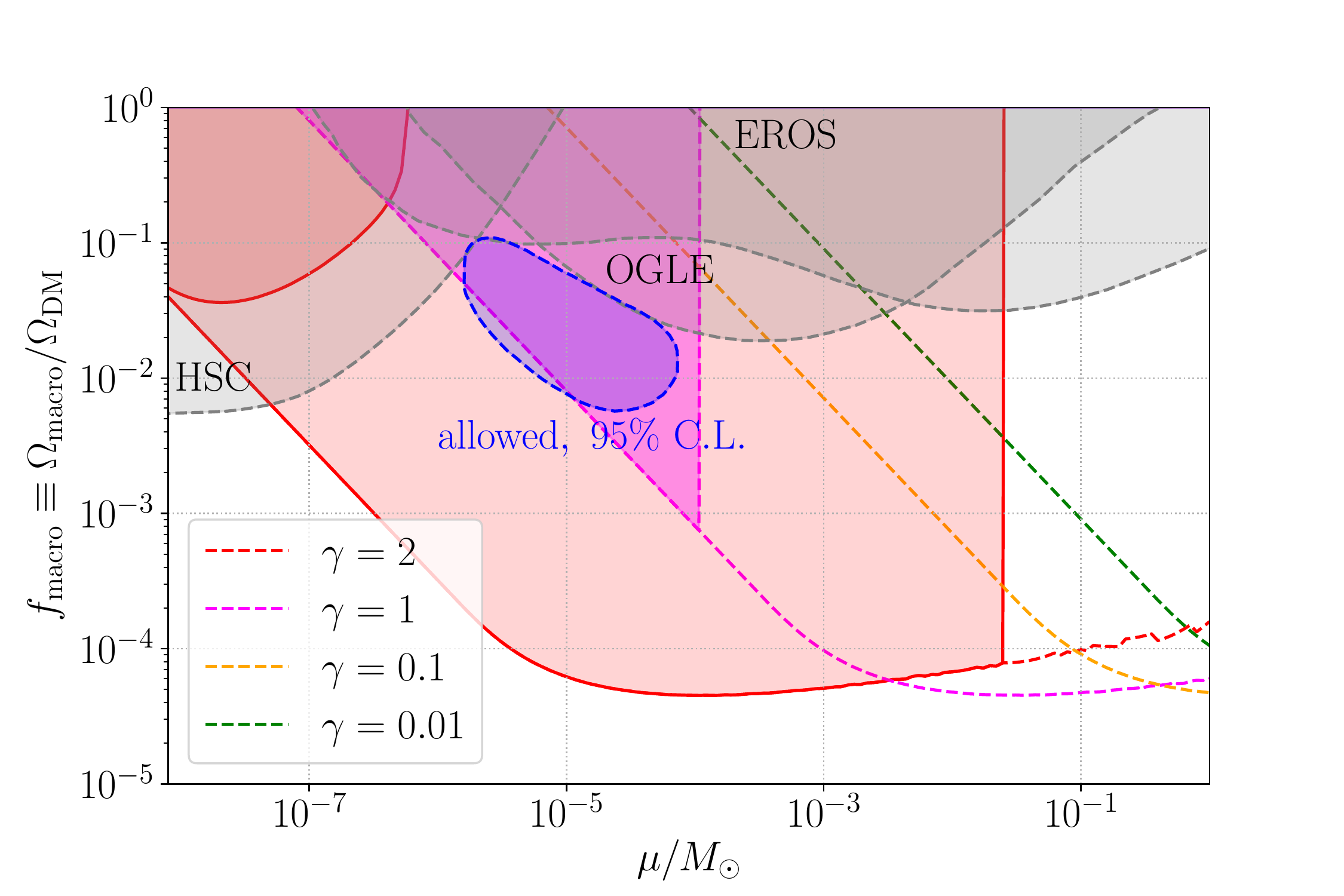}
	\caption{
	Minimum value of dark-matter fraction $f_{\rm macro}$ that LISA can detect 
		as a function of macro mass $\mu$.
		The upwards-bending branch of the $\gamma = 2$ (red, solid) curve
		uses $\tau_{\rm min} = H_{0}^{-1}$, 
		while the straight lines take $\tau_{\rm min} = 5\.{\rm years}$.
		The dark-matter distribution
		is the same as in Fig.~\ref{fig:rho}.
		The solid lines and filled regions show the parameter space where the signal is well-described
		as a gravitational-wave background. Dotted lines show an extrapolation of the stochastic 
		sensitivity in a regime where individual sources may be resolvable.
		In grey, we show microlensing constraints from 
		HSC/Subaru \cite{Niikura:2017zjd},
		EROS \cite{Tisserand:2006zx} and OGLE \cite{Wyrzykowski:2009ep, 2017Natur.548..183M}.
		The blue region displays the recently reported positive detection of ultra-short
		time-scale events attributable to planetary-mass objects 
		between $10^{-6}$ and $10^{-4}\.M_{\odot}$ \cite{Niikura:2019kqi}.
		These would contribute about $1\%$ of the dark matter, 
		which is more than expected for free-floating planets \cite{2019A&A...624A.120V}.
		 }
	\label{fig:constraints}
\end{figure}

We have performed some basic self-consistency checks on the model. First, we have estimated the total mass accreted by Sgr\.${\rm A}^{\!*}$ over the course of a Hubble time. We conservatively assume that after time $t$, all macros with $\nu / \dot{\nu} < t$ fall into the black hole. With this assumption, we find that the accreted mass is less than 10\% of the total mass of Sgr\.${\rm A}^{\!*}$ over the whole range of parameter space considered here. Second, to ensure that the cloud produces a stochastic gravitational-wave background (as opposed to a collection of resolvable sources), we require that the number of sources in the frequency band containing 90\% of the detectable signal-to-noise ratio, which we denote $\Delta f_{90\%}$, is equal to the number of frequency bins in that band, $N_{\rm bins} = T_{\rm obs} \Delta f_{90\%}$. There is a possibility of detecting individually resolvable macros for smaller spike parameters and larger macro masses when this condition is not met; we leave this for future work. In Figures~\ref{fig:SNR} and~\ref{fig:constraints}, we show the constraints that can be achieved using a stochastic search in the region. In dotted lines we show the region where the stochastic signal-to-noise ratio is consistent, even though the signal is not expected to be stochastic. Since typically a stochastic search is less sensitive than a modelled search, in the parameter space enclosed by the dotted lines, it may be possible to detect the resolvable signal from individual macros.

%%%%%%%%%%%%%%
% Stochastic SNR
%%%%%%%%%%%%%%
Bayesian data analysis methods have been developed to study the stochastic background of LISA \cite{Adams-Cornish}, including subtraction of the white-dwarf foreground \cite{Adams:2013qma}. In the context of constraining phase-transition models of the early Universe, Ref.~\cite{Caprini:2015zlo} estimated that an SNR of 10 corresponded to a detection assuming a detector with six one-directional laser links between the spacecraft. We use this same detection threshold here. The signal-to-noise ratio (SNR) of the traditional stochastic (cross-correlation) search can be written as \cite{pi-curve}
\begin{align}
	{\rm SNR}
		&=
					\sqrt{\.T\.
					\int_{0}^{\nu_{\rm max}} \d \nu\;
					\frac{ \Rcal^{2}( \nu )\.S_{\rm h}^{2}( \nu ) }
					{ P^{2}( \nu ) }
					\;}
					\label{eq:SNR}
					\; ,
\end{align}
where $\Rcal( \nu )$ is the overlap reduction function for LISA (see Ref.~\cite{Adams-Cornish}), $S_{\rm h}( \nu )$ is the strain power spectrum of the signal, which is related to $\Omega_{\rm GW}( \nu )$ by (see Ref.~\cite{pi-curve})
\begin{align}
	S_{\rm h}( \nu )
		&=
					\frac{ 3\.H_{0}^{2} }{ 2\pi^{2} }\.
					\frac{ \Omega_{\rm GW}( \nu ) }{\nu^{3}}
					\label{eq:Sh-1}
					\; ,
\end{align}
and $P( \nu )$ is the detector-power spectral density (PSD). We use the parameterization given in \cite{lisa-proposal}
\begin{align}
	P( \nu )
		&=
					\frac{ 1 }{ L^{2} }
					\left[
						S_{\rm x}( \nu )
						+
						\frac{ 1 }{ 4\pi^{2}\nu^{2} }\.
						S_{\rm a}( \nu )
					\right]
					,
\end{align}
where the arm length is $L = 2.5 \times 10^{6}\,{\rm km}$, and where the acceleration noise $S_{\rm a}$ and position noise $S_{\rm x}$ spectra are given by
\begin{align}
	S_{\rm a}( \nu )^{1/2}
		&=
					3 \times 10^{-15}\;\frac{ {\rm m} }{ {\rm s}^{2}\.\sqrt{ {\rm Hz}\,} }\,
					\sqrt{
						1
						+
						\left(
							\frac{ 0.4\.{\rm mHz}}
							{ \nu }
						\right)^{2}
					\,}
					\notag
					\\[1mm]
		&\phantom{=\;}
					\times
					\sqrt{
						1
						+
						\left(
							\frac{ \nu }
							{ 8\.{\rm mHz} }
						\right)^{4}
					\,}
					\notag
					\\[1.5mm]
	S_{\rm x}( \nu )^{1 / 2}
		&=
					10^{-11}\.
					\frac{ {\rm m} }
					{ \sqrt{\rm Hz\.}}\;
					\sqrt{
						1
						+
						\left(
							\frac{ 2\.{\rm mHz} }{ \nu }
						\right)^{4}
					\,}
					\;.
\end{align}
The collective signal-to-noise integral, Eq.~\eqref{eq:SNR}, can be computed numerically. As above, we will use $M = M_{{\rm Sgr\.{\rm A}^{\!*}}} \approx 4 \times 10^{6}\.M_{\odot}$, and the LISA PSD above. We will use $\nu_{\rm min} = 10^{-6}\.{\rm Hz}$ as the lower cutoff of the LISA band, although lowering this by an order of magnitude does not affect the results. These are depicted in Fig.~\ref{fig:SNR}. One interesting feature is that the signal is dominated by orbits close to the central black hole. This opens the possibility of directly probing the dark-matter distribution very close to the galactic centre.

In Fig.~\ref{fig:constraints} we show the minimum value of $f_{\rm macro}$ that LISA can detect as a function of macro mass, using a signal-to-noise threshold of $10$. We also show a comparison of the detection prospects with LISA, derived in this work, with microlensing constraints in the same mass range. The blue region corresponds to the recently reported positive detection of ultra-short time-scale events attributable to planetary-mass objects between $10^{-6}$ and $10^{-4}\.M_{\odot}$ \cite{Niikura:2019kqi}. These would contribute about $1\%$ of the dark matter, which is more than expected for free-floating planets \cite{2019A&A...624A.120V}. Carr {\it et al.} \cite{Carr:2019kxo} have recently shown that these observations could be naturally explained by enhancements of the PBH production through the thermal history of the Universe. Depending on the concrete spike model, LISA may provide the best detection prospects of sub-solar macroscopic dark-matter particles such as primordial black holes.\\[-2mm]

%%%%%%%%%%%%%%
% Conclusion
%%%%%%%%%%%%%%
We therefore conclude that LISA has the potential to be a robust detector of primordial-black hole dark-matter candidates. The same holds true for other macroscopic dark-matter candidates of approximately nuclear or higher density. The precise utility depends sensitively on the rate at which the orbits nearest to the black hole are replenished, and on the details of the dark-matter spike (if any). 
% Future work
In this work we have aimed at a conservative estimate of the gravitational-wave emission. There are several approximations we have made which can be improved in future work. First, due to the shape of the LISA power spectrum, the SNR is dominated by frequencies near $1\.{\rm mHz}$. In practice, this means that the signal is dominated by the orbits very close to the central black hole, which are relativistic ($v / c > 0.1$). In future work it will be interesting to explore the effect of relativistic corrections. We have also ignored effects of inclination relative to the observer, which we expect could introduce an $\Ocal( 1 )$ factor into the final result. Ref.~\cite{Lacroix:2018zmg} has placed constraints on the power law index for the dark-matter spike using observations of stellar orbits (see also Refs.~\cite{Lacroix:2013qka, Eda:2013gg}). However, because these observations are sensitive to the behavior of the spike at distances of order ten parsecs or greater from the Galactic Centre, these observations do not rule out the possibility of a dark-matter spike that falls off faster than a simple power law at these distance scales. Finally we have ignored $N$-body interactions between the macros themselves, which may cause some orbits to plunge into the black hole. 

During the time between the arXiv and journal submission, we have had very useful correspondence with anonymous referees. We have extended and improved the manuscript in several ways: we have performed a consistency check on the lifetime of the cloud, improved the discussion on other dark matter constraints, and estimated the parameter space when the signal can be described as a stochastic background. Additionally, we note that there have been several papers on related topics that have appeared. We review the contributions of Refs.~\cite{Cardoso:2019rou,Wang:2019kzb,Kavanagh:2020cfn}. Ref.~\cite{Cardoso:2019rou} considers the effect of dynamical friction on the evolution of compact binaries, and computes that the phasing of the binary. Ref.~\cite{Wang:2019kzb} computes the stochastic background from a primordial black hole cloud surrounding Sgr\.${\rm A}^{\!*}$ and the isotropic component from extragalactic sources. They find that including the effects of GW dissipation leads to an enhancement in the density of particles around the cloud. Ref.~\cite{Kavanagh:2020cfn} considers the effect of energy dissipation via dynamical friction, studying the effect on the dynamics of the cloud and computing the resulting dephasing in the gravitational waveform.

%%%%%%%%%%%%%%%%%%%%%%%%%%%%%%%%%%%%%%%%%%%%%%%%%%%%%%%%%%%%

\begin{acknowledgements}
We indebted to Vitor Cardoso, Enrico Barausse, Emanuele Berti, and Paolo Pani for invaluable remarks on the evaluation of the characteristic gravitational-wave strain amplitude, correcting a mistake in the previous version of this manuscript. We also thank Joe Bramante, Gil Holder, Marc Kamionkowski, Marco Peloso, and Monica Valluri for helpful comments and useful discussions. G.D.S.~thanks the Oskar Klein Centre Cosmoparticle Physics for their hospitality and F.K.~thanks Case Western Reserve University and Lawrence Berkeley National Laboratory for their hospitality while this work was completed. K.F.~and F.K.~acknowledge support from DoE grant DE-SC0007859 at the University of Michigan as well as support from the Leinweber Centre for Theoretical Physics. K.F.~and F.K.~acknowledge support by the Vetenskapsr{\aa}det (Swedish Research Council) through contract No.~638-2013-8993 and the Oskar Klein Centre for Cosmoparticle Physics. G.D.S.~is partially supported by Case Western Reserve University grant DOE-SC0009946. 
\end{acknowledgements}

%%%%%%%%%%%%%%%%%%%%%%%%%%%%%%%%%%%%%%%%%%%%%%%%%%%%%%%%%%%%

\bibliographystyle{spphys}
\bibliography{refs}

\end{document}